**Superstripes and complexity in high-temperature superconductors**

Antonio Bianconi[1,2] and Nicola Poccia[1]

[1]*Department of Physics, Sapienza University of Rome, P. le A. Moro 2, 00185 Roma, Italy*
[2]*Rome International Center for Materials Science (RoCMat) Superstripes, Via dei Sabelli 119A, 00186 Roma, Italy*

*Abstract:*
While for many years the lattice, electronic and magnetic complexity of high-temperature superconductors (HTS) has been considered responsible for hindering the search of the mechanism of HTS now the complexity of HTS is proposed to be essential for the quantum mechanism raising the superconducting critical temperature. The complexity is shown by the lattice heterogeneous architecture: a) heterostructures at atomic limit; b) electronic heterogeneity: multiple components in the normal phase; c) superconducting heterogeneity: multiple superconducting gaps in different points of the real space and of the momentum space. The complex phase separation forms an unconventional granular superconductor in a landscape of nanoscale superconducting striped droplets which is called the "superstripes" scenario. The interplay and competition between magnetic orbital charge and lattice fluctuations seems to be essential for the quantum mechanism that suppresses thermal decoherence effects at an optimum inhomogeneity.

\**web site*: www.rocmat.eu   *E-mail:* antonio.bianconi@superstripes.net

**1. Introduction.**

The search for new materials with a higher superconducting critical temperature has been carried out during the last 100 years. In the early period, the material research was focusing on pure and simple pure metals made by a single element. At ambient pressure niobium has the highest superconducting critical temperature 9.3 K. The research later shifted to more complex materials: binary alloys. The highest critical temperature was reached in 1973 in a binary compound $Nb_3Ge$ ($T_c$ = 23 K). In 1986, high-temperature superconductivity ($T_c$ = 41 K) was found in perovskites made of ternary $La_2CuO_{4+y}$ and quaternary $La_{2-x}Ba_xCuO_{4+y}$ compounds [1,2]. The barrier of liquid nitrogen temperature ($T_c$ = 77 K) was overcome in 1987 in a quaternary compound $YBa_2Cu_3O_{6+y}$. The critical temperature is found to increase in a pentanary compound $HgBa_2Ca_2Cu_3O_{8+y}$ with the number of elements in the chemical formula and the maximum critical temperature known so far 139 K has been reached in 1994 in $Hg_{0.2}Tl_{0.8}Ba_2Ca_2Cu_3O_{8.33}$. The increasing chemical complexity indicates that the maximum T$_c$ emerges in a unique particular metallic phase with low symmetry and with a very particular complex Fermi surface selected between a very large numbers of possible cases. Indeed the high external pressure pushes metallic lithium (with a T$_c$ lower than 4 mK at zero pressure) towards low symmetry structures and its superconducting transition goes up to 23 K at 50 GPa. In cuprates T$_c$ reaches the record of 164 K in HgBa$_2$Ca$_2$Cu$_3$O$_{12\,y}$ under 30 GPa pressure. The pressure pushes the system near structural phase transitions where the lattice becomes soft with nanoscale lattice phase separation and the chemical potential is pushed near a Lifshitz critical point in the electronic multiband structure where electronic phase separation occurs. These facts show that the high-temperature superconductivity (HTS) occurs in very particular heterogeneous phases forming a new class of granular superconductors.

BCS theory was developed in 1957. In the standard approximation for uniform three-dimensional metals with high carrier density and the pairing interaction is considered homogeneous in the real space and in the momentum space. The high critical temperature is expected for a high density of states at the Fermi level, a high coupling term and a large Debey frequency. On the contrary the experimental high T$_c$ research of Alex Muller addressed 3D perovskite ceramics with low carrier density and multiple electronic bands in particular on the interplay of two electronic states with different orbital symmetry $3d_{x^2-y^2}$ and $3d_{z^2}$ near the Fermi level where Jahn-Teller electron-lattice interaction breaking the Born–Oppenheimer approximation could give bipolarons [2]. On the contrary the dominant theoretical approach was based on the search for the mechanism of HTS in a strongly correlated two-dimensional electron gas near a Mott-insulator at half filling [3]. In 1993 it was proposed that high T$_c$ occurs in "particular lattice architectures", a layered heterostructure at the atomic limit, and at "particular" charge carrier density. [4,5]. The lattice nanoscale architecture a "superlattice of quantum units" gives quantum size effects forming multiband metals with bands of different symmetry where single electron interband scattering is forbidden by selection rules. The chemical potential is tuned near a band edge i.e., near a Lifshitz critical point of a 2.5 Lifshitz electronic phase transition where quantum resonance between open and closed scattering channels known as Fano resonance (or "shape resonance", or "Feshbach resonance") in a multi-gap superconductor occurs [6]. The Fano resonance is reached by fine tuning the chemical potential in a multiband metal near the Lifshitz critical points for a vanishing Fermi surface or opening a neck in a Fermi surface [7,8]. The system is near the lattice and electronic instability with coexisting free particles in a large Fermi surface and polaronic carriers in a small Fermi surface [5-8] that show up in a complex phase granular superconductor made of striped superconducting droplets called "superstripes" [9,10].

This proposal was confirmed in 2001 when the record of the superconducting critical temperature of 39 K in binary intermetallics was found in magnesium diboride made of a heterostructure at the atomic limit formed by boron atomic layers intercalated by magnesium atomic layers [11-13]. The lattice is close to structural instability and the chemical potential is near a Lifshitz critical point for opening a neck in the small sigma Fermi surface. In 2008, the discovery of a new family of ternary iron-arsenide-based superconductors [14,15] has confirmed that the layered heterostructure at the atomic limit and the control of the misfit strain between different modules [14] are essential features for emerging high $T_c$. The efforts to increase the complexity of these materials have been rewarded by a maximum superconducting critical temperature of ~ 50 K.

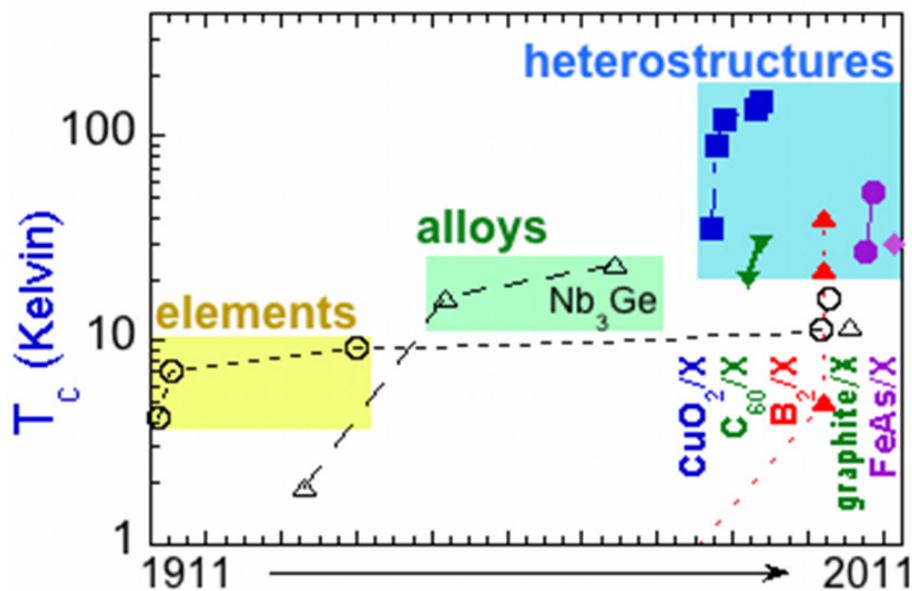

**Figure 1.** The superconducting critical temperature in different families of metals as a function of time. The first group on the left side of the plot includes pure elements showing superconductivity. The second group, spanning about 50 years of material research from 30's to 80's, includes alloys showing superconductivity. The third group, spanning these last 25 years of material research, includes composites made of up to 6 different elements, made of heterostructures at the atomic limit like $CuO_2$ or boron, or graphene or FeAs monolayers, intercalated by spacer layers X, that show the highest critical temperature known so far. Also a heterostructure made of quantum dots, like $C_{60}$, with spacer ions show high $T_c$ superconductivity.

Pressure studies have shown the possibility of high critical temperature in iron-selenium-based multilayers [16]. This shows how high pressure investigations are at the core of research for new functional materials to approach lattice instabilities, that are needed for high-temperature superconductivity. For example in a binary compound iron-selenium made of a heterostructure at the atomic limit the superconducting critical temperature could be raised through doping, (e.g., potassium), achieving a superconducting critical temperature of ~31 K [17].

The history of materials research for high $T_c$ superconductors, briefly explained so far, is synthetically reported in Fig.1. The data points have been divided in three ensembles: pure elements, alloys and heterostructures. It is now well established that all high-temperature superconductors are layered heterostructures at the atomic limit. In Fig.2 we report the maximum critical temperature reached in different compounds. In pure metallic elements the maximum critical temperature is 9.2 K for Nb and 23 K in Li under pressure. In binary compounds the maximum temperature is 39 K in $MgB_2$. Between the other binary compounds with high critical temperature it is relevant to remark that also doped fullerene compounds and $Nb_3Ge$ are made of heterostructures at atomic limit : a 3D superlattice of quantum dots and a superlattice of quantum wires respectively. The maximum critical temperature in a ternary compound has been reached in $La_2CuO_{4+y}$ reaching about 45 K. The maximum critical temperature in a ternary compound has been found in $La_2CuO_{4+y}$ reaching about 45 K. The maximum critical temperature in a quaternary compound has been found in $YBa_2Cu_3O_{6+y}$ reaching about 85 K. The maximum critical temperature in a pentanary compound has been found in $HgBa_2Ca_2Cu_3O_{8+y}$ reaching about 130 K. The record of the critical temperature at ambient pressure has been found in $Hg_{0.2}Tl_{0.8}Ba_2Ca_2Cu_3O_{8+y}$ made of six different elements reaching about 139 K. Therefore the maximum critical temperature occurs in unique complex structure, and increasing the number of different constituting atomic elements is needed to find the very particular material where robust high-temperature superconductivity appears.

In the materials shown in Fig. 2, complexity appears both in momentum and real space. The role of disorder and complexity has been recognized as a fundamental ingredient for achieving better superconducting performance in the heterostructure at the atomic limit after 2005 [18-21]. Intriguingly, the relevance of the proximity to a Lifshitz transition in a multi-gap superconductor [8] is now emerging clearly in iron-based superconductors [22-29].

Design of heterostructures at atomic limit by superlattice of quantum wires have been theoretically proposed [8], however, the field of the control of critical temperature by manipulation of the lattice structure has advanced slowly. Advances in the precision of molecular beam epitaxy and pulsed laser deposition have allowed to better design and characterize interfaces [30], because of the significant improvements in the control of epitaxial strain (depending on the thickness of the spacers) and of the misfit strain (depending on the bond length of the layered heterostructure at the atomic limit) [31]. In cuprates, in fact, it has been shown that the superconducting critical temperature is controlled by the misfit strain and a three variables phase diagram is needed to unify all cuprate materials showing the dome of hill at a particular doping and misfit strain [31-33]. Theoretically, dipolar insulators and polarized interfaces in layered heterostructures at the atomic limit have been recently considered as a good candidate for high critical superconducting temperatures [34,35]. On the other hand, direct control of metastable interfaces at the atomic limit is experimentally challenging. Even if the interest in perovskites is moving from bulk ceramics and single crystals to thin films, multilayers and superlattices, new synchrotron radiation techniques are unveiling for the first time the real space organization of superlattices in single crystals with an unprecedented level of details [36,37], providing inspiration for theorists [36].

It is known that interstitial dopant atoms roam at the interfaces and also if they are fundamental for the establishing of the complex state adapted for high-temperature superconductivity, they introduce a degree of disorder very detrimental for applications and for more complex hierarchic heterostructures. This is one of the main reason why for example in manganites scanning tunneling techniques are difficult to apply [39]. Generally speaking, alloys with mobile ions in some range of temperature or strain are know in the ionic conductors are of large interest in the chemistry of catalitic device [40]. Here, however, the concept of ionic conductors is mixed with that of heterostructures at atomic limit, leading to a high degree of complexity in the materials as well as in the theory of high-temperature superconductivity. Interstitial dopant atoms have been known for a long time to be crucial also in cuprates, however, resolving the real space phase separation of ordered oxygen interstitials from disordered regions has been considered experimentally highly demanding. This apparent drawback for applications should be good for manipulation and design of interfaces. The powerful X-rays, in fact, can be used to change the properties of an oxide superconductor, thus effectively writing superconducting regions within an insulating matrix [41].

According to a growing number of experimental results [42,43], the future of functional materials and high-temperature superconductivity is linked to the control of the complexity of the interfaces and superlattices in the heterostructures at the atomic limit. Coupling of several lattice instabilities at the interfaces have been shown in multi-ferroic materials to raise special properties to room temperature [44]. Lattice instabilities, however, do not belong only to heterostructures at the atomic limit and some particular alloys (e.g., $Nb_3Ge$) are in the vicinity of a lattice instability [45]. The design of such systems with multi-component instabilities (magnetic, lattice, orbitals) and the precise controlled relaxation of such instabilities by chemical dopants and/or external perturbation is going to attract interest in the coming years [46-48, 41]. Relaxation of the instabilities could be the cause of self-organization of the scale-free superlattices at the atomic limit or the creation of new ones with commensurate or incommensurate order with spatial broken symmetries and non-trivial correlations as recently shown in high-temperature cuprate superconductors [36,37,48]. In this paper, we have followed this thought, dividing the different aspects of complexity of the momentum and real space in high-temperature superconductors in six points: multi-component; multi-gaps; structural complexity; transport; spin fluctuations and pairing; polarons.

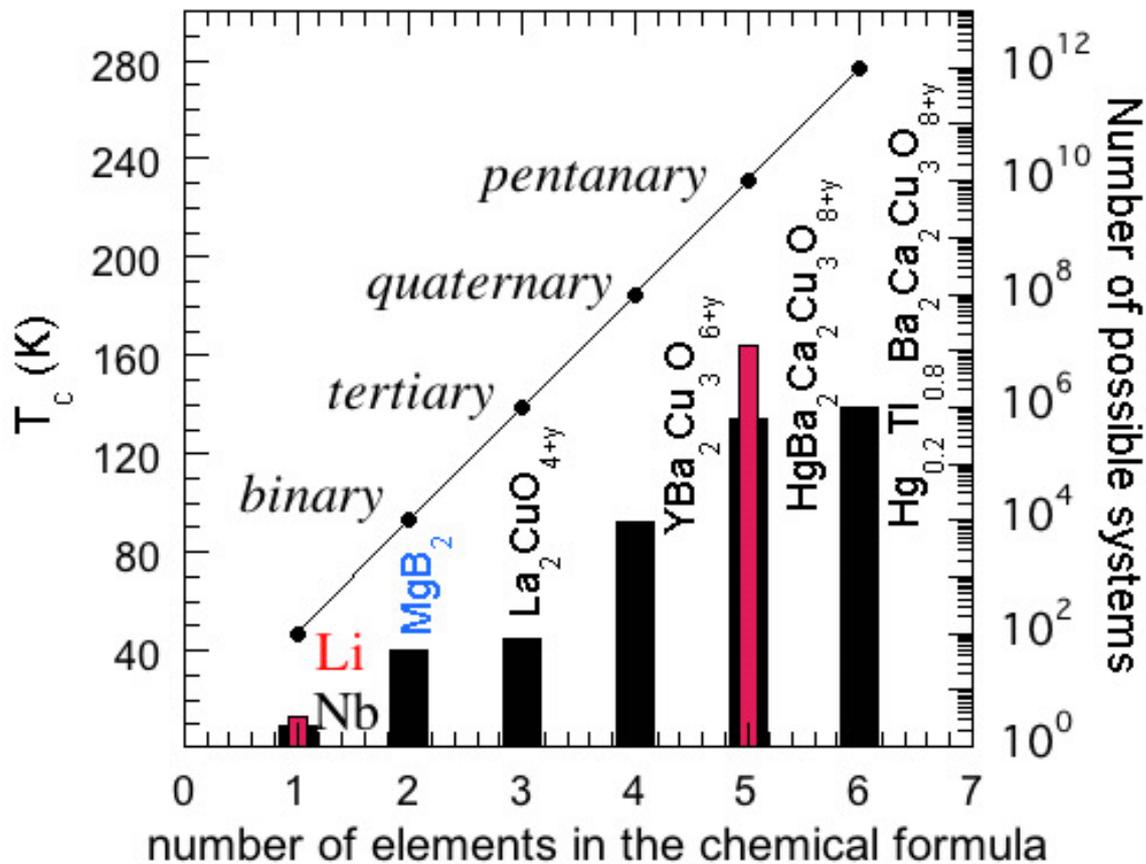

**Figure 2**: The maximum superconducting critical temperature reached by material research in systems made by a single element, two elements (binary compounds), three elements (ternary compounds), four elements (ternary compounds), five elements (pentanary compounds) and six elements. All high Tc superconductors are camposite matrials made of multiple structural modules of units and their electronic structure has multiple Fermi surface and multiple gaps in the superconducting phase

# 1. Multi-component.

The specific feature of cuprates superconductor is the fact that the doped holes L forms a new set of states $3d^9\underline{L}$ at the insulator-metal transition. High resolution, polarized X-ray absorption near edge structure (XANES) study of the structure and symmetry of the unoccupied hole states in $YBa_2Cu_3O_{7-y}$ by varying y has identified an extra spectral density from 932 to 935.5 eV for y=0.0-0.15 which appears only at the insulator-metal transition at y = 0.3. This new peak is due to transitions from $3d^9\underline{L}^*$ to $\underline{2p}3d^{10}\underline{L}^*$ final state, where $\underline{L}^*$ denotes the hole in the ligand orbital (present in the ground state) [50]. The first evidence of multi-band superconductivity in cuprates has been found by XANES in $YBa_2Cu_3O_{7-y}$ for y>0.66 where the $3d^9\underline{L}$ states in the $CuO_2$ plane coexist with the Cu $3d^9\underline{L}$ states in the $CuO_2$ chains [51].

This experiment has provided not only evidence for multi-band superconductivity, but also for the spatial segregation of two sets of electronic states in two different portions of the same crystal.

The coexistence of polarons associated with local lattice distortions in nanoscale domains and free carriers in cuprates has been observed in $La_{2-x}Sr_xCuO_4$ derived from susceptibility measurements by Müller et al. [52] and by EXAFS experiments [53].

Mitsen et al. have proposed a model where a pair of $3d^9L$ states is located in two adjacent cations forming a two-atom negative-U center (NUC). Owing to virtual transitions of electron pairs to these NUCs, states $(\overline{k},\uparrow -\overline{k}\downarrow)$ in the vicinity of the FS are pair-wise coupled, which leads to superconducting pairing in the system. In fact, the localized charges are a source for local closing of cation-anion gap favoring the realization of an unusual electron-electron interaction paring mechanism. The superconducting scenario depend on the microstructure of different types of dopant ion ordering that is discussed in the frame of doped $La_2CuO_4$ [54]. This model, that has been extended by Ivanenko et al. to discuss the superconductive pairing mechanism in pnictides, is based on the coexistence of localized pairs and free carriers [55].

Maier has used dynamic cluster quantum Monte Carlo simulations of multi-band Hubbard models to describe many of the complex phenomena, observed in the cuprates and iron-based high-temperature superconductors [56]. The simplest case for multi-band superconductivity in cuprates is based on multiple Fermi surfaces in a bilayer Hubbard model with inter-layer hopping. Maier shows that charge stripes in the 2D Hubbard model can lead to a significant enhancement of superconductivity [57]. In the results presented both the t-J and Hubbard models have spin-fluctuation pairing glue.

Theoretically, emphasis on multiband superconductivity in a high-$T_c$ cuprate, iron based and aromatic superconductors providing, has been proposed by Aoki [58]. The relationship between the inhomogeneous orbital distribution and the superconductivity has been investigated by Wakisaka, et al. considering α-FeSe$_{1-x}$Te$_x$ [59]. The authors have studied the electronic structure of FeSe$_{1-x}$Te$_x$ in comparison with Ir$_{1-x}$Pt$_x$Te$_2$ discussing the relationship between their orbital states and their superconductivities. ARPES measurements on FeSe$_{1-x}$Te$_x$ suggest that the Fe 3d yz/zx orbital degeneracy at Γ point and orbitally induced Peierls [60] effect in the tetragonal lattice play key roles for superconductivity.

## 2. Multi-gaps.

Multi-gap superconductivity was proposed in the early times of BCS superconductivity as an extension of the standard single band BCS theory. The experimental evidence for multi-gap superconductivity has been provided 44 years later for the first time in MgB$_2$ because it was assumed that even a very small impurity scattering will drive superconductors to the dirty limit, averaging the k-dependence of the superconducting gaps. A two-band model of MgB$_2$ superconductivity has been developed by Kristoffel et al. It exploits repulsive σ-π interband coupling besides the σ-intraband electron-phonon attraction and Coulomb interaction [61].
Multi-gap superconductivity model for the description of cuprate superconductor characteristics on doping scale (hole and electron) has been based on the leading interband pairing channel that couples an itinerant band and defect states created by doping [62]. Recently, Ord et al. have applied the Ginzburg-Landau equations for a two-band superconductor, derived from the Bogoliubov-de Gennes equations and the relevant self-consistency conditions for a two-gap system [63]. The authors find critical and non-critical coherence lengths in the spatial behavior of the fluctuations of order parameter. The dependence of the rigid coherence length and the microscopic length have been shown as a function of the interband interaction. The non-monotonic temperature dependence of coherence lengths appear in the superconducting phase if the interband coupling is sufficiently weak. The interband pairing can give Fano resonances near the Lifshitz critical point in multiband superconductors with mini-bands formed by quantum size effects [64,65] where quantum fluctuations are suppressed by the 3D arrays of quantum units [. Superconducting instabilities in 3D and 2D extended the Hubbard model with Coulomb repulsion between electrons on neighboring sites in the limit of low electron density on simple cubic (square) lattice have been investigated by Kagan et al. [66] In this model the influence of electron-polaron effects and other mechanisms for mass-enhancement (related to the momentum dependence of the self-energies) on the effective mass and scattering times of light and heavy components in the clean case (electron-electron scattering and no impurities) has been analyzed. A tendency towards phase separation (towards negative partial compressibility of heavy particles) has been found [67].

Recently, Kagan et al. have focused on a two-band Hubbard model with one narrow band considering the limit of low electron densities in the bands and strong intraband and interband Hubbard interactions [68]. The most compelling evidence for multi-gap superconductivity has been provided by iron-based superconductors.

The experiments show multiple gaps using directional point-contact Andreev-reflection (PCAR) measurements on high-quality single crystals of the e-doped 122 compound $BaFe_{1.8}Co_{0.2}As_2$. The existence of two superconducting gaps with no line nodes on the FS, and whose amplitude is almost the same in the ab plane or along the c-axis has been shown [69]. Recently, Tortello et al. performed PCAR measurements in 122 and 1111 electron-doped pnictides compounds [70].

The Dynes-ansatz procedure has been proposed by Garcia [71] to extract the superconductive energy gap from the tunneling differential conductance measured by STM. They propose a generalized model that makes possible to study thermal and quantum fluctuations in STM experiments. This technique has been used to investigate evolution of superconductivity as a function of temperature and particle size in single, isolated Pb nanoparticles [72].

## 3. Structural Complexity

The multi-scale phase separation from atomic scale to nano-scale and from mesoscopic scale to micron-scale is now emerging as an essential feature for establishing high-temperature superconductivity. Solving numerically the Cahn-Hilliard (CH) equation, it is possible to follow the time evolution of a coarse-grained order parameter which satisfies a Ginzburg-Landau free-energy functional commonly used to model superconductors has shown by De Mello et al. [73]. In this model, the zero resistivity transition arises when the intergrain Josephson coupling $E_J$ is of the order of the thermal energy and phase locking takes place among the superconducting grains [74]. This approach has explained the pseudogap and superconducting phases and it also reproduces some recent STM data [75]. Recently, De Mello et al. have performed calculations by the CH theory, in order to describe the phase separation of oxygen interstitial in the spacer layers separating the $CuO_2$ planes in $La_2CuO_{4+y}$ [76].

The possibility that the strain field associated with the structural supermodulation in Bi-2212 and certain other cuprate materials may modulate the superconducting pairing interaction has been discussed by Andersen et al [77]. The pinning of electronic stripes to structural distortions has been discussed by Andersen et al. [78] using self-consistent mean field calculations to solve the Bogolibov-de Gennes equations on a square lattice, near a van Hove singularity.

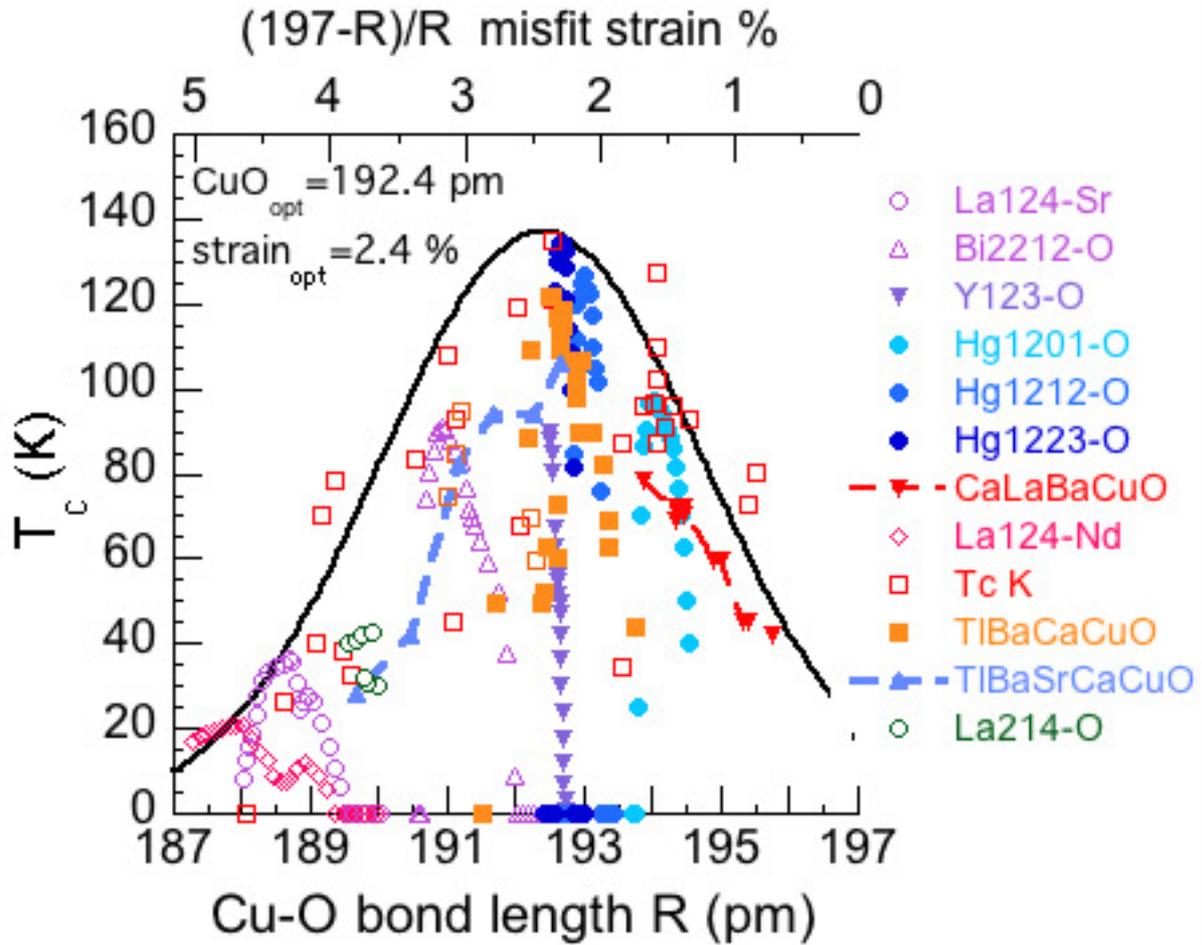

**Figure 3**. The critical temperature for all cuprate superconductors plotted as a function of the Cu-O bond length. The critical temperature at a fixed average Cu-O bond length is controlled by changing the charge transfer between active and spacer layers and decreased by lattice disorder, however at optimum charge transfer and lower disorder the maximum Tc is controlled by the lattice misfit strain. The optimum distance is 192.4 pm and the undistorted fully relaxed Cu-O bond is 197±1 pm, therefore the optimum misfit strain, that gives the optimum inhomogeneity is 2.5%

Multiple non-magnetic impurities on the local electronic structure of the high-temperature superconductor of Zn doped Bi2212 have been investigated by STM by Machida et al. They have found several fingerprints of quantum interference of the impurity bound states including a two-dimensional modulation of local density-of-states with a period of approximately 5.4 A along the

a- and b-axes and an abrupt spatial variations of the impurity bound state energy [79]. Recently using TM, Nakamura et al have shown that in single crystals of Zn doped $Bi_2Sr_{1.6}La_{0.4}CuO_{6\ \delta}$ the number of the resonances was less than that of the doped impurities (0.45% determined by inductively coupled plasma optical emission spectrometry) [80].

The striped hole self-organization in cuprate $CuO_2$ layers has been proposed by Mitin in the framework of a string model implying the direct overlap of oxygen orbitals with the mean hole density [81]. Recently, Mitin gives support to the important role played by the crystal structure symmetries in determining the electronic properties of complex material such as layered high-temperature superconductors. Mitin has performed simulations of the holes trajectories of hopping hole pairs in $CuO_2$ fragments of $La_2CuO_{4\ \delta}$ and $HgBa_2Ca_3Cu_3O_{9\ \delta}$ in order to describe the phase separation of oxygen interstitial in the spacer layers separating the $CuO_2$ planes in $La_2CuO_{4\ y}$ [82]. The results indicate an anisotropic structural inhomogeneity along the *c*-axis that could be related with the modulation of the bonding between layers. Recently, Ricci et al. have proposed a new heterostructure at the atomic limit, thanks to the novel insight in the structure of complex materials which are intrinsically disordered [83-87].

## 4. Transport

Glassy features have been found in in $La_{1.97}Sr_{0.03}CuO_4$ by c-axis magnetotransport mesurements and resistance noise study. The in-plane magnetoresistive of the single crystal below the spin glass transition temperature, becomes positive and exhibits several glassy feature, such as history dependence, memory and hysteresis [88]. The authors claim that the present results strongly suggest that the memory effects reflect primarily the behavior of doped holes in the copper plane. It is known that no-linear current voltage characteristic is a sign of the transition from superconductivity to normal conduction. The results indicate that the charge dynamics become increasingly slow and correlated as temperature T→0 [89]. Recently, Raicevic et al. describe the $La_{2-x}Sr_xCuO_4$ the temperature dependence of the zero-field cooled in plane resistance below 30 K, the in-plane resistance as a function of time upon the application of a variable magnetic field as well as the difference between different cooling protocols [90].

To create a wide range of different defects and to tailor the electrical and superconducting properties in high-temperature superconductors, Lang et al. have developed a technique based on ion irradiation. Depending on the species of ions used during the irradiation, their energy and fluence, nanoscale columnar pinning centers can be created that enhance the critical current, or randomly distributed point defects that change the superconducting properties [91,92].

Recently, Lang et al have measured the resistivity and Hall conductivity of optimum doped and underdoped YBCO as a function of the corrected temperature in a magnetic field [93].

In optimally doped samples this non-ohmic effect is limited to temperatures below 100 K. In a moderately underdoped sample, however, the non-linearity extends to the temperature range from $T_c$ = 53 K to 150 K. Therefore the main result of the work is the non-ohmic transport behavior in YBCO that is best seen in the Hall conductivity measurements. In the presence of perturbations that break translational symmetry, such as impurities and defects, the Fermi liquid readjusts itself, producing a spatially inhomogeneous pseudopotential "seen" by quasiparticles. The formation of ''ripples,'' the Friedel oscillations, surrounding the perturbation have been discussed in the singular case of two-dimensional systems [94], Recently, Andrade et al. have investigated theoretically the effects of weak disorder scattering in a correlated host. Hartree-Fock and dynamical field theory. Calculations have been used to study the scattering process of quasiparticles off the screened disorder potential. They have shown that both the local and non-local contributions to the renormalized disorder potential are suppressed in strongly renormalized Fermi liquids [95].

Low system dimensionality has been shown to play a relevant role in the superconductor-to-normal transition in superconducting nanowires [96]. In particular, low dimensionality in an organic layered superconductor $(TMTSF)_2ClO_4$ is associated with the upturn of the curve of the upper critical field at low temperatures [97]. Croitoru et al. have shown the anisotropy of the in-plane critical field in uniform and non-uniform phases of layered superconductors within the quasi-classical approach, taking into account the interlayer Josephson coupling. The analysis of the amplitude and the field direction dependence of the onset temperature of the superconductivity is explained as the interplay of the system dimensionality, orbital pair-breaking, Pauli paramagnetism, and a possible formation of the FFLO state.

## 5. Spin Fluctuations and Pairing

The pairing symmetry of the multiorbital pnictide superconductor $BaFe_{2-x}Co_xAs_2$ has been investigated by Raman scattering by Sugai et al [98]. Recently, Sugai et al. describe the magnetic excitations in $La_{2-x}Sr_xCuO_4$ of the separated dispersion induced by the superlattice structure of the stripes in distinction from the excitations by two-magnon Raman scattering [99].

The authors measure the dispersion of the magnetic excitations in spin-charge stripes with different holes concentrations and compare the results with the model of Seibold et al. The electronic inhomogeneities in underdoped cuprates have been investigated by Seibold et al, focusing on the material dependent textures in the framework of a three bands model. Uniform magnetic spirals have been found at small doping which are unstable towards nanoscale phase separation. The authors propose a comparison with one-band calculations which have revealed stable spirals in the overdoped regime for a large ration of the hopping terms among the 4s copper orbitals [99].

Thalmeier et al. investigate the frustrated two-dimensional $S=1/2$ next nearest neighbor Heisenberg antiferromagnet on a square lattice [100]. The elementary spin wave excitations of the model have been obtained by transforming to a local coordinate system aligned with the moment orientation and by performing a Holstein Primakoff approximation. Staggered moment of intermediate anisotropic model as a function of frustration ration has been shown. This model is particularly relevant for layered $S=1/2$ vanadium oxides.

It is known that the magnetic field can be used to probe the doping and momentum dependence of the superconductor gap. Huang et al shows the change in the penetration depth as a function of temperature and magnetic field. The calculations have been done starting from the t-J model on a square lattice, taking into account the 2D geometry of cuprate superconductors [101].

## 6. Polarons

Alexandrov investigates strong electron-lattice interaction as the glue for high-temperature superconductivity and describes the analytical multi-polaron theory in the strong coupling regime for doped highly polarizable ionic insulators [102]. The theory of dense polaronic systems in the intermediate coupling regime still remains poorly understood. Weak and intermediate Coulomb repulsions do not result in the Cooper instability in p- or d-wave channels in 3D and 2D with or without lattice- induced band-structure effects. Based on some variational simulations the authors argue that the repulsive hard-sphere (i.e., Hubbard U) model does not account for high-temperature superconductivity in the strong-coupling regime either [103].

Starting from a generic Hamiltonian including unscreened Coulomb and Frohlich interactions, the author shows the possibility to express the bare Coulomb repulsion and electron phonon interaction through material parameters rather than computing them from first principles in many physically important cases. The many particle electron system is then described by an analytically solvable polaronic t-Jp Hamiltonian with reduced hopping integral. A high-temperature superconducting state is proposed to be driven by small super-light bipolarons protected from clustering [33].

It is known that variation in the Cu-O bond stretching in cuprates, dramatically affects the high critical temperature properties of the materials. De Filippis et al have studied theoretically the optical conductivity of cuprates in the low-density limit of the t-t'-J model taking into account the hole-lattice coupling. The authors also predicted the influence of the isotope substitution on the optical conductivity whose fine details depend on whether or not the material contains apical oxygen [104]. Recently, Cataudella et al. have developed a t-t'-J-Holstein in a two-dimensional lattice geometry to calculate the bond stretching phonon mode [105].

The phonon spectral function for different parameters of the model show a phonon softening, observed in experiments on underdoped cuprates. The authors propose that such experimental results could be explained as the result of the interaction among strongly correlated holes and the anti-ferromagnetic fluctuations without invoking strong charge inhomogeneities.

Hori et al. have investigated superconductive pairing mechanisms in a two-orbital Hubbard model coupled with Jahn-Teller phonons on a 2D lattice [106]. A practical realization is the $Na_xCoO_2 \cdot yH_2O$ ($x \approx 0.35$, $y \approx 1.3$) which is a superconductor with a $T_c$ of about 5 K and consists of two-dimensional $CoO_2$ layers separated by a thick insulating layer of Na ions and $H_2O$ molecules [107].

The authors show that the electron-phonon and electron-electron interactions collaborate inducing a novel pairing state characterized by spin singlet, orbital singlet, and odd parity in k-space.

Marsiglio addresses the general issue of polaron formation in the Holstein model in one, two and three dimensional lattices [108]. The problem is tackled within perturbative and exact approaches showing that the polaronic crossover becomes sharper upon decreasing the phonon frequency, but never becomes a sharp transition as long as the phonon frequency stays finite.

This result is well known, but it is interestingly confirmed within the technical frameworks adopted here. In the one dimensional case, the electron has a polaronic character for any coupling strength. This result seems at odds with previous numerical exact diagonalization results, where the polaronic crossover was found at finite coupling even on one-dimensional lattices as shown by Capone et al. [109] This discrepancy arises from the fact that the exact diagonalization calculations were carried out on rather small lattices, where the one-dimensional van Hove singularity at the bottom of the band is cut off by the finite spacing in momentum space (in some sense this would make the calculation with one electron similar to the case of a lattice with finite density of electrons).

This does not seem the case in the present calculation and might originate the polaronic character in 1D down to very small couplings [110].

Vidmar et al have used the time-dependent t-J Holstein model to describe the nonequilibrium dynamics of the Holstein polaron driven by an external electric field. In the strong-coupling limit, weakly damped Bloch oscillations, consistent with nearly adiabatic evolution within the polaron band, persist up to extremely large electric fields. The shape of the traveling polaron has been investigated in detail [111]. Recently, the authors have extended this work for the non-equilibrium scenario in strongly correlated materials [112].

**Conclusion**

The scenario of granular superconductivity proposed for cuprates [113] is emerging as a common feature of high-temperature superconductors. While the superconducting nanoscale granular units show a striped pattern the grains can form arrays as in complex networks with multiple components [114]. It has recently shown that an optimum inhomogeneity with an increasing second moment of the distribution of superconducting grains can drive the critical temperature for the insulator to superconductor phase transition toward high-temperature [115]. The cuprates hold the record of the highest critical temperature (160 K) [11] occurring at a critical value of the Cu-O bond length. In Figure 3 we report the critical temperature as a function of the Cu-O bond length for all cuprate superconductors at different charge transfer and different disorder that induce a variation of the critical temperature from zero to maximum $T_c$..

The Cu-O bond length has measured in selected samples by a local probe Extended X-ray Absorption Structure (EXAFS) (51) able to identify different polymorphic structures in the same crystal (116). The maximum $T_c$ as a function the Cu-O bond length clearly shows a Gaussian like behavior centered at the Cu-O bond length, 192.5 pm.

The variation of the Cu-O bond length at constant optimum doping and minimum disorder is due to the misfit strain between the ultra-narrow $CuO_2$ atomic layer and the spacer layers of these lamellar materials. The optimum interatomic distance for highest $T_c$ at ambient pressure is about 2.5% smaller than the equilibrium distance of the Cu-O bond length, 197 pm, [118] measured for the $Cu^{2+}$ ion forming a $CuO_4$ plaquette in solution by X-ray Absorption Near Edge Structure (XANES) a method that probes the high order inter-atomic correlation function and therefore able to detect different coordination geometries in oxides [119]. Therefore in this phase diagram for all cuprate perovskite families [116] the highest critical temperature occurs not only for an optimum doping but also for a optimum compression of the $CuO_2$ plane due to the misfit strain with the spacer layers.

The efforts are now addressed to understand the role of organization of dopants ions in the spacer layers [41] and the corrugation of the $CuO_2$ lattice due to the misfit strain [49]. This lattice disorder is the basis for the appearance of a unique granular electronic phase that shows high $T_c$ granular superconductivity in a "quantum-critical state", where the electrons form collective patterns that look the same regardless of scale forming heterogeneous networks of linked superconducting grains with scale-free distribution [114]. These materials show therefore multiphase complexity (structural, magnetic and electronic) that has hindered physicists search for the mechanism of high $T_c$ for many years since the observed phenomenology depends on the time and spatial sensitivity of the experimental probes. The insulator-to-superconductor transition can therefore be controlled in these complex materials by hydrostatic pressure, irradiation by x-ray, and strong electric fields at the interfaces opening many venues to new materials and new technologies.